\documentclass[11pt, a4paper]{article}
\usepackage[a4paper, margin=2.5cm]{geometry}
\usepackage{amsmath,amssymb}
\usepackage{hyperref}
\usepackage{graphicx}
\usepackage{booktabs}
\usepackage{authblk}

\title{\textbf{Semantically Disentangled Attention for Multi-Camera Road Segmentation under Missing Camera Conditions}}

\author{Vijay John\thanks{\texttt{vjohn@ltu.edu}}}
\author{Chan-Jin Chung\thanks{\texttt{cchung@ltu.edu}}}
\affil{Department of Mathematics and Computer Science, Lawrence Technological University}

\date{}

\begin{document}

\maketitle

\begin{abstract}
\noindent Road surface segmentation is an important perception task for
autonomous racing that requires robust perception with low
latency and memory requirements.
To achieve robust semantic segmentation accuracy, multiple
cameras are used within a multi-camera sensor fusion
formulation. However, existing multi-camera methods assume that both
cameras are always available during training and inference.
This assumption is not guaranteed in autonomous racing, as
camera failure due to adverse environmental or mechanical
conditions is possible, resulting in the
\textit{missing camera problem}. To address this, we propose a novel multi-camera road
segmentation framework comprising a learned null image
substitution module, a semantically disentangled attention
module, dual ASPP decoder heads, and an inter-camera label
consistency module.
To account for varying sensor configurations including single
missing camera data, a two-stage curriculum learning strategy
is adopted, where the first stage learns inter-camera
representations from fully-available data.
The trained framework is then fine-tuned and adapted to
single missing camera data using configuration-aware losses
and camera presence flags.
A comparative analysis on the RoRaTrack autonomous racing
dataset demonstrates that the proposed framework consistently
improves average available camera Road~IoU and substantially
improves average missing camera Road~IoU over baseline
algorithms.
\end{abstract}

\section{Introduction}
\label{sec:intro}
\begin{figure}[t]
\centering
\includegraphics[width=\textwidth]{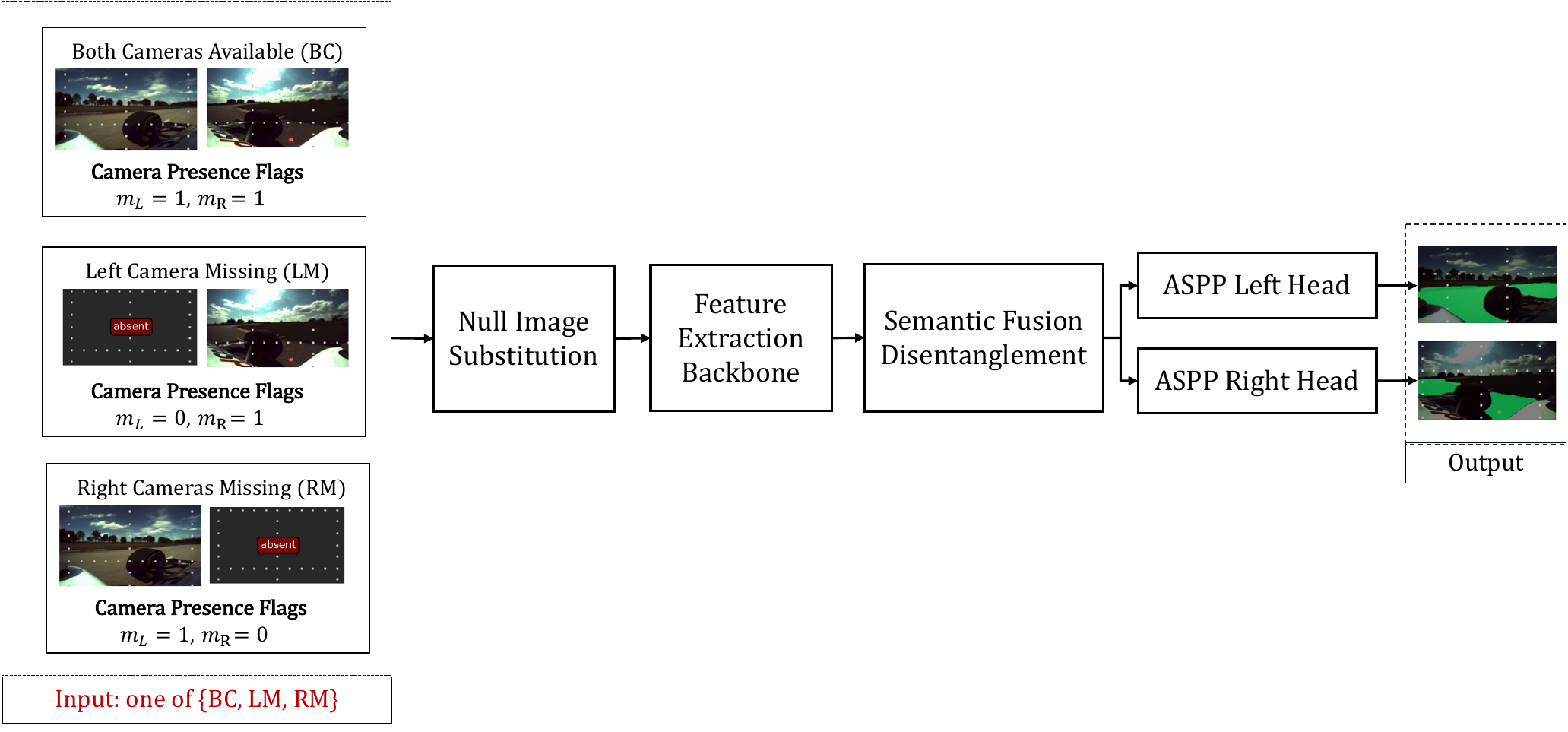}
\caption{Overview of the proposed framework. Input from one of three sensor configurations is processed through different modules to produce left and right road segmentation predictions.}
\label{fig:overview}
\end{figure}

Road surface segmentation for autonomous racing is an important safety-critical perception problem that requires robust perception at low latency.
Road surface segmentation identifies the pixel-level free space, which along with the obstacles detected feeds into the path planning module. Inaccurate road surface segmentation can result in the racing vehicle either missing the racing line or leaving the track entirely. 

Racing vehicles are typically equipped with multimodal sensors such as the visible cameras, LiDAR, and radar, to aid in robust perception~\cite{Betz_2022,Betz_2023}. Among these, visible-camera-based perception frameworks exploit rich appearance information for free-space estimation and obstacle detection~\cite{Chen_2018_ECCV, Cordts_2016}. Multi-camera-based perception frameworks leverage multi-view data for more robust perception~\cite{Li_BEVFormer,chitta2022transfuserimitationtransformerbasedsensor}. However, these methods rely on the availability of all sensor data during both training and inference. But in real-world environments such as in autonomous racing, an individual sensor failure is a possibility. In such a scenario, the perception algorithm should still reliably perform perception accounting for the missing modality. This problem is referred to as the \textit{missing modality
problem} in the literature~\cite{JohnTOMM2025}, and in this
work the missing modality is specifically a single camera
in a dual-camera configuration.

For autonomous racing, the operating environments introduces camera-specific failure conditions including lens contamination, image saturation due to direct sun glare, and vibration-induced misalignment. These conditions can cause individual camera failure and severely degrade road segmentation performance. Existing approaches for the missing modality problem are formulated for heterogeneous sensor pairs such as camera and Lidar~\cite{havaei2016hemisheteromodalimagesegmentation,Hussen2020,recasens2023zorro}. Consequently, these approaches do not exploit the geometrical and spatial information present inherently in co-located camera-based perception. 

In this work, we propose a novel multi-camera road segmentation framework for autonomous racing addressing the missing modality problem while exploiting the inter-camera geometrical and spatial information. The proposed framework, termed RoadFormer-SDA, performs
multi-camera road segmentation through a learned null image
substitution, an inter-camera label consistency (ILC) module, and a Semantically Disentangled
Attention~(SDA) module. The ILC module enforces
geometric consistency between left and right segmentation
predictions through a learned dense flow field during
training. The SDA module contains  multiple intra-camera self-attention and inter-camera cross-attention mechanisms
for geometric correspondence, photometry alignment, and
extrinsic calibration. An overview of the proposed framework is presented in Fig.~\ref{fig:overview}.

The framework is trained with varying sensor configurations,
including both cameras available and single missing camera
configurations, using a two-stage curriculum learning
strategy.
In the first stage, the framework learns inter-camera
representations from fully-available data.
In the second stage, the trained framework is fine-tuned
using both available and missing camera configurations
through config-aware loss masking.

A comparative analysis on the RoRaTrack autonomous racing
dataset~\cite{ghosh2025racingdatasetbaselinemodel} with baseline algorithms validates the
proposed framework, reporting consistent improvement in
average available camera Road~IoU and substantially improved average missing camera Road~IoU over baseline algorithms.

The novel contributions of this work to the literature are
as follows:

\begin{itemize}
\item \textbf{Semantically Disentangled Attention~(SDA):}
  A novel multi-camera fusion module with five independent
  attention mechanisms, with inter-camera cross-attention and intra-camera self-attention.

\item \textbf{Inter-camera label consistency~(ILC) module:}
  A novel auxiliary training module that enforces geometric
  consistency between left and right segmentation predictions
  through a learned dense flow field.

\item \textbf{Two-stage curriculum learning:}
  A principled training curriculum that separates inter-camera representation learning on fully-available data from missing camera data through config-aware loss masking.
\end{itemize}

The remainder of this paper is organized as follows.
Section~\ref{sec:related} reviews related work.
Section~\ref{sec:method} presents the proposed framework
architecture.
Section~\ref{sec:training} details the two-stage training
procedure and loss functions.
Section~\ref{sec:experiments} reports experimental results,
comparative analysis, and ablation study.
Section~\ref{sec:conclusion} concludes.

\section{Related Work}
\label{sec:related}

Semantic segmentation of road surfaces is a fundamental perception task
in autonomous driving, requiring pixel-wise classification of the driveable surface under diverse real-world conditions. Significant advances have been made in perception for autonomous driving in urban environment using large scale datasets~\cite{Cordts_2016,Neuhold_2017}. However, perception algorithms formulated for general urban driving are not suitable for high-speed autonomous racing. Betz et al.~\cite{Betz_2022} surveyed autonomous vehicle racing across
perception, planning and control, identifying the unique challenges posed
by high-speed autonomous racing. In the context of the Indy Autonomous Challenge, Betz
et al.~\cite{Betz_2023} presented the TUM autonomous racing pipeline, demonstrating the low latency and memory requirements for full-scale autonomous racing.

The literature on semantic segmentation for autonomous driving in urban environments reports state-of-the-art performance using deep learning frameworks based on the encoder-decoder and transformer architectures.  Long et al.~\cite{Long_FCN} established the encoder-decoder paradigm with the fully convolutional network.
Chen et al.~\cite{Chen_2018_ECCV} proposed DeepLabv3+, extending this
with separable convolution and a dedicated decoder module to
refine segmentation results along object boundaries. Wang et al.~\cite{wang2021_PAMI} proposed HRNet, which maintains high-resolution representations throughout the network via parallel high-to-low resolution images. Subsequently,  Xie et al.~\cite{Xie_2021_Segformer} proposed SegFormer, unifying
transformers with lightweight MLP decoders into a simple segmentation framework.
Mehta and Rastegari~\cite{Mehta_2022_MobileViT} proposed MobileViT,
which uses element-wise self-attention to achieve
high-accuracy segmentation under the memory and latency constraints of edge devices.

While single-modality approaches report state-of-the-art performance, they are limited by adverse environmental conditions and individual sensor limitations. To enhance the robustness of vision-based perception in various adverse environmental conditions, multi-modal fusion-based perception frameworks have been proposed by combining complementary information across sensor images.
Valada et al.~\cite{valada16iser} demonstrated that fusion
of RGB, near-infrared and depth data outperforms single-modality approaches for semantic scene understanding. With the introduction of the transformer architecture with attention mechanism by Vaswani et al.~\cite{vaswani2017attention}, several researchers have utilized transformer to significantly advance the perception accuracy for tasks such as object detection~\cite{CarionDETR} and semantic segmentation~\cite{Li_BEVFormer}. Li et al.~\cite{Li_BEVFormer} proposed the BEVFormer, aggregating features
from multiple cameras into a bird's-eye-view representation
using transformers with cross-attention mechanism, reporting state-of-the-art perception accuracy. Chitta et al.~\cite{chitta2022transfuserimitationtransformerbasedsensor} proposed TransFuser, fusing image and LiDAR representations through self-attention at multiple resolutions for autonomous driving waypoint prediction.

While these methods demonstrate effective multi-sensor fusion, they
assume full sensor availability at both training and inference time,
an assumption that cannot be guaranteed in autonomous racing. The missing modality problem arises when one or more sensor streams are unavailable at training or inference time. In recent years, researchers have identified this problem and have proposed solutions for perception tasks across different domains including medical imaging~\cite{havaei2016hemisheteromodalimagesegmentation} and indoor person recognition~\cite{JohnTOMM2025}. Havaei et al.~\cite{havaei2016hemisheteromodalimagesegmentation} proposed
HeMIS, for medical imaging, learning a modality-specific embedding into a shared latent space for each MRI sequence, such that available modalities at inference time
can be averaged over without imputing or synthesizing the missing data to produce the final
segmentation. Xiong et al.~\cite{Xiong_2025} extended this with a guided-learning network that trains specialized models for different missing modality
scenarios using dual learning paths to capture both multimodal and
modality-specific representations. Hussen Abdelaziz et al.~\cite{Hussen2020} applied a modality dropout approach, where input streams are stochastically removed during training. Recasens et al.~\cite{recasens2023zorro} proposed Zorro, using masks to
control how inputs from each modality are guided inside transformers,
keeping some of the representation modality-pure when a stream is
absent. John and Kawanishi~\cite{john2022multimodal,john2023audiovisual,john2023multimodal,JohnTOMM2025}, address this problem using metric learning and cascaded learning frameworks for indoor-based perception tasks such as audio-visual or audio-visual-thermal person recognition. A detailed survey of the deep multimodal learning under missing modalities is presented in the work of Wu et al.~\cite{wu2026deepmultimodallearningmissing}.

Beyond architectural design, training a robust perception
framework under missing modality conditions requires an analysis of data complexity. Jointly training on fully-available and missing modality
configurations is not straightforward, as modules that
depend on all modalities being present for training are
degraded by missing modality configurations.
Curriculum learning~\cite{bengio_curriculum_2009}, proposed
to improve model convergence by training in a structured
order from simple to complex data, can be applied here to
separate fully-available modality training from missing
modality adaptation.

While the above methods address the missing modality problem for heterogeneous sensor pairs such as RGB-LiDAR or
audio-visual inputs, multi-camera perception is a
homogeneous sensor pair setting with its own challenges. Firstly, like other multi-modal frameworks, existing multi-camera fusion methods also assume full sensor availability~\cite{Li_BEVFormer,chitta2022transfuserimitationtransformerbasedsensor}.
An additional challenge in multi-camera road segmentation
is class imbalance, where the driveable surface constitutes
a minority of pixels compared to non-driveable pixels.
Lin et al.~\cite{lin2017focal} proposed Focal Loss to
concentrate learning on hard minority-class examples to
address this imbalance. Apart from class imbalance, loss functions can also be used to learn inter-camera relationships. In literature, view synthesis
and photometric consistency losses have been used in
multi-camera frameworks to learn depth and ego-motion from
monocular
video~\cite{zhou2017unsupervised,godard2019monodepth2},
while cross-view geometric consistency losses have been
used to model viewpoint-induced camera variations~\cite{zhao2026crossviewdomaingeneralizationgeometric}. 

\noindent\textbf{Research Gap.}
Existing missing modality
methods focus on heterogeneous sensor pairs and are not designed for multi-camera frameworks~\cite{havaei2016hemisheteromodalimagesegmentation, Xiong_2025,Hussen2020}. Moreover, existing multi-camera fusion
methods~\cite{Li_BEVFormer,chitta2022transfuserimitationtransformerbasedsensor}
assume all camera availability, leaving the challenge of
multi-camera road segmentation under partial sensor
availability in autonomous racing unaddressed.
To address this, we propose RoadFormer-SDA, which combines
semantically disentangled attention, learnable null
representations, and a two-stage training curriculum for
robust multi-camera road segmentation under both full and
partial sensor availability.

\section{Proposed Framework}
\label{sec:method}

\begin{figure}[t]
\centering
\includegraphics[width=\textwidth]{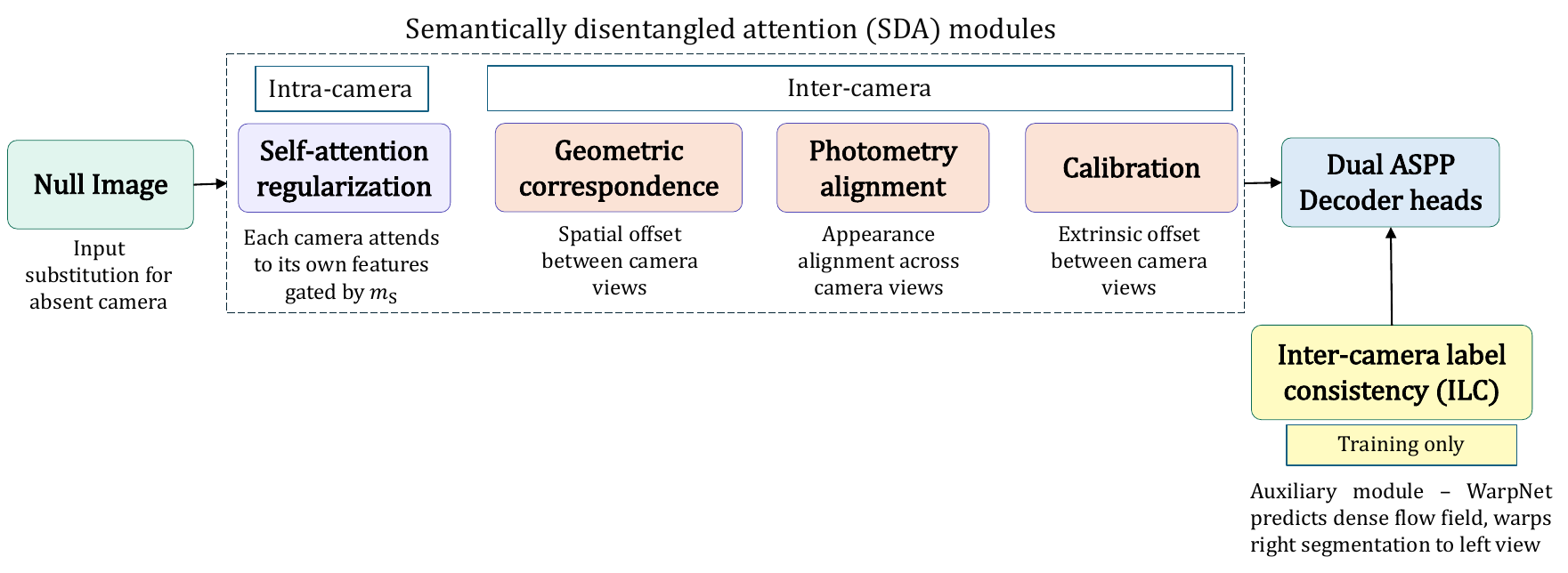}
\caption{Proposed framework modules: null image substitution,
semantically disentangled attention~(SDA), dual ASPP decoder
heads, and inter-camera label consistency~(ILC) auxiliary
training module.}
\label{fig:sda}
\end{figure}

In this paper, we propose RoadFormer-SDA, a multi-camera road segmentation framework under full and partial
sensor availability.
Let $\mathbf{I}_L, \mathbf{I}_R \in \mathbb{R}^{3 \times H \times W}$
denote the left and right camera images respectively, and let
$m_L, m_R \in \{0, 1\}$ denote the corresponding camera presence flags,
where $m_s = 1$ indicates camera $s \in \{L, R\}$ is available and
$m_s = 0$ indicates camera $s$ has failed. Given the input images and camera presence flags, RoadFormer-SDA
estimates road segmentation maps for both camera views,
$\hat{y}_L, \hat{y}_R \in \{0, 1\}^{H \times W}$, under
varying sensor configurations: both cameras present~(BC,
$m_L = m_R = 1$), left camera missing~(LM, $m_L = 0,\,
m_R = 1$), and right camera missing~(RM, $m_L = 1,\,
m_R = 0$).

The proposed framework consists of a sequential pipeline comprising an input null images module, a feature extraction backbone, a semantically disentangled attention~(SDA) module, dual ASPP decoder heads, and an inter-camera label consistency~(ILC) head.

Given the inputs $\mathbf{I}_L$, $\mathbf{I}_R$ and presence flags
$(m_L, m_R)$, the input null images module first operates on the raw
camera inputs, substituting a learned pixel-level null representation
$\mathbf{N}_s$ for $\mathbf{I}_s$ when camera $s$ is absent,
producing conditioned inputs $\tilde{\mathbf{I}}_L$ and
$\tilde{\mathbf{I}}_R$.
Next, the feature extraction backbone extracts feature maps
from each conditioned input independently, producing feature maps
$\mathbf{F}_L$ and $\mathbf{F}_R$.
Following feature extraction, the SDA module uses multiple attention heads to obtain the inter-camera and intra-camera representations from the two camera images. This given as input to the dual ASPP decoder heads which 
produce $\hat{y}_L$ and $\hat{y}_R$ for each view independently. To obtain a robust segmentation output, an ILC head is used to enforce consistency between $\hat{y}_L$ and $\hat{y}_R$ during training.

The camera presence flag $m_s \in \{0, 1\}$ is provided per sample for both training and testing, and plays an important role in the curriculum learning strategy. An overview of the proposed architecture is shown in Figure~\ref{fig:sda}, with the details being presented next.

\subsection{Input Null Images and Feature Extraction Modules} Given the inputs $\mathbf{I}_L$, $\mathbf{I}_R$ and camera presence flags $(m_L, m_R)$, this module obtains a learned pixel-level null representation $\mathbf{N}_s \in \mathbb{R}^{3 \times H \times W}$ for $\mathbf{I}_s$ when camera $s$ is absent. This results in conditioned inputs $\tilde{\mathbf{I}}_L$ and $\tilde{\mathbf{I}}_R$, given as,

\begin{equation}
  \tilde{\mathbf{I}}_s = m_s \cdot \mathbf{I}_s +
  (1 - m_s) \cdot \mathbf{N}_s
  \label{eq:null_image}
\end{equation}

$\mathbf{N}_s$ is initialized to the ImageNet channel mean and then updated during training.

Next, the feature extraction backbone extracts feature maps from each conditioned input $\tilde{\mathbf{I}}_s$ independently, resulting in feature map $\mathbf{F}_L$ and $\mathbf{F}_R$. Since $\mathbf{N}_s$ is initialized and updated around the ImageNet channel mean, the feature extraction backbone is based on an ImageNet pre-trained model.
 
\subsection{Semantically Disentangled Attention (SDA) Module}
The SDA module learns intra-camera and inter-camera
representations from $\mathbf{F}_L$ and $\mathbf{F}_R$,
where $\mathbf{F}_s \in \mathbb{R}^{C \times H_f \times W_f}$
with $C$ denoting the feature channel dimension, through five
independent attention mechanisms. This includes two intra-camera
self-attention mechanisms and three inter-camera
cross-attention mechanisms for geometric correspondence,
photometry alignment, and calibration.

Prior to the attention operations, spatial average pooling
is applied to $\mathbf{F}_s$ to reduce token count, producing
positional-encoded token sequences $\mathbf{t}_s \in
\mathbb{R}^{N_p \times C}$, where $N_p = \frac{H_f}{2} \times
\frac{W_f}{2}$.

Each attention mechanism computes independent query, key,
and value representations from $\mathbf{t}_s$ using learned
weight projections of dimension $d = C/4$.

The term \textit{semantically disentangled} refers to the
assignment of a dedicated loss function to each attention
mechanism, where each loss supervises a specific
inter-camera relationship. An overview of the modules is presented in Fig.~\ref{fig:sda}.

\subsubsection{Self-Attention Mechanism}
The two intra-camera self-attention mechanisms capture
spatially relevant features within each camera image
independently:
\begin{equation}
  \mathbf{A}^{\mathrm{self}}_s =
  \mathrm{Attn}(Q^{\mathrm{self}}(\mathbf{t}_s),\,
                K^{\mathrm{self}}(\mathbf{t}_s),\,
                V^{\mathrm{self}}(\mathbf{t}_s)) \cdot m_s
  \label{eq:self_head}
\end{equation}
where $m_s$ gates the full attention output to zero when
camera $s$ is absent.

\subsubsection{Cross-Attention Mechanisms}
The three inter-camera cross-attention mechanisms are each
computed bidirectionally, with each camera stream querying
the opposite camera's features.
Each direction is gated independently by the camera presence flag of the key-value stream. More specifically, $\mathbf{A}_{LR}$, where the left camera
tokens provides the query and the right camera tokens
provides the key-value, is gated by $m_R$. Similarly for
$\mathbf{A}_{RL}$, where the right camera tokens provides
the query and the left camera tokens provides the
key-value, is gated by $m_L$.

\paragraph{Geometry Mechanism}
Since the two cameras are mounted on the vehicle with
known relative positions, the spatial offset between their
views is a learnable geometric relationship. The geometry mechanism is given as:
\begin{equation}
  \mathbf{A}^{\mathrm{geo}}_{LR} =
  \mathrm{Attn}(Q^{\mathrm{geo}}(\mathbf{t}_L),\,
                K^{\mathrm{geo}}(\mathbf{t}_R),\,
                V^{\mathrm{geo}}(\mathbf{t}_R)) \cdot m_R
  \label{eq:geo_head}
\end{equation}
with $\mathbf{A}^{\mathrm{geo}}_{RL}$ computed with the
query and key-value tokens swapped, gated by $m_L$.

\paragraph{Calibration Mechanism}
Although the two cameras are rigidly mounted on the vehicle,
the extrinsic transformation between them is subject to
drift due to vibration and mechanical stress associated with autonomous racing, making it a learnable relationship.
The calibration mechanism uses shared QKV projections across
both streams and is given as:
\begin{equation}
  \mathbf{A}^{\mathrm{calib}}_{LR} =
  \mathrm{Attn}(Q^{\mathrm{calib}}(\mathbf{t}_L),\,
                K^{\mathrm{calib}}(\mathbf{t}_R),\,
                V^{\mathrm{calib}}(\mathbf{t}_R)) \cdot m_R
  \label{eq:calib_head}
\end{equation}
with $\mathbf{A}^{\mathrm{calib}}_{RL}$ computed with the
query and key-value tokens swapped, gated by $m_L$.

\paragraph{Photometry Mechanism}
Although both cameras are subject to the same global ambient lighting conditions, they exhibit view-specific
photometric variations due to their different positions and
angles. Thus, the appearance alignment between their images is learnable. The photometry mechanism is formulated to learn this alignment. The photometry mechanism is given as:
\begin{equation}
  \mathbf{A}^{\mathrm{photo}}_{LR} =
  \mathrm{Attn}(Q^{\mathrm{photo}}(\mathbf{t}_L),\,
                K^{\mathrm{photo}}(\mathbf{t}_R),\,
                V^{\mathrm{photo}}(\mathbf{t}_R)) \cdot m_R
  \label{eq:photo_head}
\end{equation}
with $\mathbf{A}^{\mathrm{photo}}_{RL}$ computed with the
query and key-value tokens swapped, gated by $m_L$.

\subsubsection{Attention Fusion}
Following the attention mechanisms, for each camera
$s \in \{L, R\}$, the self-attention output and three
cross-attention outputs, each of dimension $d$, are
concatenated to dimension $4d = C$ and projected with a
residual connection to $\mathbf{t}_s$, followed by a
feed-forward network:
\begin{equation}
  \mathbf{z}_s = \mathrm{FFN}\!\left(
    \mathrm{Proj}\!\left(
      \mathbf{A}^{\mathrm{self}}_s \,\|\,
      \mathbf{A}^{\mathrm{geo}}_{s} \,\|\,
      \mathbf{A}^{\mathrm{photo}}_{s} \,\|\,
      \mathbf{A}^{\mathrm{calib}}_{s}
    \right) + \mathbf{t}_s
  \right)
  \label{eq:fused_stream}
\end{equation}
where $\mathbf{A}^{\mathrm{geo}}_{s}$,
$\mathbf{A}^{\mathrm{photo}}_{s}$, and
$\mathbf{A}^{\mathrm{calib}}_{s}$ denote the cross-attention output for camera $s$ querying the opposite camera.

\subsection{Dual ASPP Decoder Heads}
Using the output of fusion, the dual ASPP decoder heads produce road segmentation logits
from $\mathbf{F}^{\mathrm{fused}}$.
Two independent ASPP heads~\cite{Chen_2018_ECCV}, each
utilizing dilated convolutions at rates $\{1, 6, 12\}$ and
global average pooling, produce $\hat{y}_L$ and $\hat{y}_R$
respectively.

\subsubsection{Inter-Camera Label Consistency (ILC) Module}
The ILC module enforces geometric consistency between the
predicted segmentation maps $\hat{y}_L$ and $\hat{y}_R$
during training.
A lightweight convolutional warp
network~\cite{jaderberg2015spatial} takes the left and right
logits as input and predicts a 2D dense flow field $\phi
\in \mathbb{R}^{H \times W \times 2}$:
\begin{equation}
  \phi = \mathrm{WarpNet}([\hat{y}_L,\, \hat{y}_R])
  \label{eq:warp_net}
\end{equation}
Following flow field prediction, a warped left-view
prediction $\hat{y}^w_L$ is constructed by displacing each
pixel coordinate of the $H \times W$ output space by $\phi$
and sampling the corresponding logit values from $\hat{y}_R$
via bilinear interpolation~\cite{jaderberg2015spatial}.  The ILC module is used during training only and is removed
at inference.

\subsection{Loss Functions}
Different attention mechanisms in the SDA module and the output heads are supervised by a
dedicated loss function that targets a specific learning
objective. The geometry, photometry, and calibration losses
supervise the inter-camera cross-attention mechanisms. The fusion loss regularizes the attention fusion. The segmentation and edge-weighted segmentation losses supervise the decoder heads, and the
ILC loss enforces geometric consistency at the prediction
level. Together, these seven losses form the total training
objective. The combined training objective is the weighted sum of all seven losses:
\begin{align}
  \mathcal{L}_{\mathrm{total}} \;=\;&
    w_1 \mathcal{L}_{\mathrm{seg}}
  + w_2 \mathcal{L}_{\mathrm{e\text{-}seg}}
  + w_3 \mathcal{L}_{\mathrm{fusion}}
  \nonumber\\
  &+ w_4 \mathcal{L}_{\mathrm{geo}}
  + w_5 \mathcal{L}_{\mathrm{photo}}
  + w_6 \mathcal{L}_{\mathrm{calib}}
  + w_7 \mathcal{L}_{\mathrm{label}}
  \label{eq:total_loss}
\end{align}
with loss weights summarized in Table~\ref{tab:loss_weights}.

The weights are chosen empirically to balance the
contribution of each loss to the total gradient and are
not normalized. The active losses for each training stage and sensor configuration are detailed in Section~\ref{sec:training}. We next explain each loss in detail.

\paragraph{Geometry Attention Loss}

The geometry loss $\mathcal{L}_{\mathrm{geo}}$ supervises
the cross-attention mechanism to learn spatial correspondence between the two camera views.

Two auxiliary learned projections $\mathbf{W}^L_{\mathrm{geo}}$
and $\mathbf{W}^R_{\mathrm{geo}}: d \rightarrow C$, used
only during training, map the attention outputs from
dimension $d$ to the full channel dimension $C$, enabling
the L1 supervision against the query-camera token sequences
$\mathbf{t}_s \in \mathbb{R}^{N_p \times C}$.
These projections are discarded at inference as they serve
no role beyond providing the geometry supervision signal
during training.

An L1 loss is then computed between each projected attention
output and the original query-camera token sequence:
\begin{equation}
  \mathcal{L}_{\mathrm{geo}} =
  \frac{1}{2}\!\left(
    \|\mathbf{W}^L_{\mathrm{geo}}\mathbf{A}^{\mathrm{geo}}_{LR} -
      \mathrm{sg}(\mathbf{t}_L)\|_1 +
    \|\mathbf{W}^R_{\mathrm{geo}}\mathbf{A}^{\mathrm{geo}}_{RL} -
      \mathrm{sg}(\mathbf{t}_R)\|_1
  \right)
  \label{eq:geo_loss}
\end{equation}
where $\mathrm{sg}(\cdot)$ denotes the stop-gradient
operator, applied to the target token sequences to prevent loss collapse by ensuring only the attention output
side receives the gradient.

In the geometry attention mechanism, the query camera tokens
attend to the opposite camera tokens and reconstruct their
own token sequence, constraining the mechanism to learn
what spatial information the query camera can recover from
the opposite view.
This models the geometric offset between the two camera
positions.

\paragraph{Calibration Attention Loss}
The calibration loss $\mathcal{L}_{\mathrm{calib}}$
supervises the calibration mechanism using two auxiliary learned projections
$\mathbf{W}_{\mathrm{fwd}}: d \rightarrow C$ and
$\mathbf{W}_{\mathrm{inv}}: d \rightarrow C$. The auxiliary projections are used only
during training, and they map the LR and RL attention outputs to the full channel dimension respectively, enabling L1 supervision against the opposite-camera token sequences
$\mathbf{t}_s \in \mathbb{R}^{N_p \times C}$.
These projections are discarded at inference as they serve
no role beyond providing the calibration supervision signal
during training.

The calibration loss $\mathcal{L}_{\mathrm{calib}}$ is given as,
\begin{equation}
  \mathcal{L}_{\mathrm{calib}} =
  \frac{1}{2}\!\left(
    \|\mathbf{W}_{\mathrm{fwd}}\mathbf{A}^{\mathrm{calib}}_{LR} -
      \mathrm{sg}(\mathbf{t}_R)\|_1 +
    \|\mathbf{W}_{\mathrm{inv}}\mathbf{A}^{\mathrm{calib}}_{RL} -
      \mathrm{sg}(\mathbf{t}_L)\|_1
  \right)
  \label{eq:calib_loss}
\end{equation}

In the calibration attention mechanism, the query camera
tokens attend to the opposite camera tokens and reconstruct
the opposite camera token sequence. This mechanism learns the extrinsic transformation between
the two camera positions bidirectionally, modeling both the
forward~($L \rightarrow R$) and inverse~($R \rightarrow L$)
directions jointly.

\paragraph{Photometry Attention Loss}
The photometry loss $\mathcal{L}_{\mathrm{photo}}$ supervises
the photometry mechanism using cosine similarity, which is
invariant to feature magnitude and focuses on directional
alignment between the two photometry representations.

Two auxiliary learned projections $\mathbf{W}^L_{\mathrm{photo}}$
and $\mathbf{W}^R_{\mathrm{photo}}: d \rightarrow C$, used
only during training, map the LR and RL attention outputs
to the full channel dimension, followed by
$\ell_2$-normalization to obtain
$\hat{\mathbf{n}}^{\mathrm{photo}}_L$ and
$\hat{\mathbf{n}}^{\mathrm{photo}}_R$ respectively.

The cosine similarity is then computed between
$\hat{\mathbf{n}}^{\mathrm{photo}}_L$ and
$\hat{\mathbf{n}}^{\mathrm{photo}}_R$, requiring both
cameras to be present and is therefore active for BC
samples only:
\begin{equation}
  \mathcal{L}_{\mathrm{photo}} = 1 -
  \frac{1}{N_p}\sum_{i=1}^{N_p}
  \hat{\mathbf{n}}^{\mathrm{photo}}_L(i) \cdot
  \hat{\mathbf{n}}^{\mathrm{photo}}_R(i)
  \label{eq:photo_loss}
\end{equation}

where $\hat{\mathbf{n}}^{\mathrm{photo}}_s \in
\mathbb{R}^{N_p \times C}$ are the $\ell_2$-normalized
photometry representations for stream $s$, and $N_p$ is
the number of tokens.

\paragraph{Attention Fusion Loss}
The attention fusion loss $\mathcal{L}_{\mathrm{fusion}}$ loosely constrains the fused per-camera representation $\mathbf{z}_s$ to remain close to the original input token sequences $\mathbf{t}_s$, ensuring the attention fusion output preserves the $s$-th camera's relevant information while containing relevant inter-camera information.

The camera identity $s$ in $\mathbf{z}_s$ corresponds to
the intra-camera self-attention output and the query
contributions from the three cross-attention mechanisms.
The loss is computed as the average L1 distance over both
cameras:
\begin{equation}
  \mathcal{L}_{\mathrm{fusion}} =
  \frac{1}{2}\!\left(
    \|\mathbf{z}_L - \mathrm{sg}(\mathbf{t}_L)\|_1 +
    \|\mathbf{z}_R - \mathrm{sg}(\mathbf{t}_R)\|_1
  \right)
  \label{eq:fusion_loss}
\end{equation}
where $\mathrm{sg}(\cdot)$ denotes the stop-gradient
operator, applied to the input token sequences to ensure
only the fused output side receives the gradient.

\paragraph{Segmentation Loss}
The segmentation losses are defined over the two decoder outputs, containing a primary
segmentation loss $\mathcal{L}_{\mathrm{seg}}$ and an
edge-weighted segmentation loss
$\mathcal{L}_{\mathrm{e\text{-}seg}}$ that emphasizes track
boundary pixels.
The primary segmentation loss $\mathcal{L}_{\mathrm{seg}}$
is a combination of cross-entropy and soft Dice losses,
computed for both camera images:
\begin{equation}
  \mathcal{L}_{\mathrm{seg}} =
  \mathcal{L}_{\mathrm{CE+Dice}}(\hat{y}_L, y_L) +
  \mathcal{L}_{\mathrm{CE+Dice}}(\hat{y}_R, y_R)
  \label{eq:seg_loss}
\end{equation}
where $\mathcal{L}_{\mathrm{CE+Dice}} =
0.6\,\mathcal{L}_{\mathrm{CE}} +
0.4\,(1 - \mathrm{Dice}_{\mathrm{soft}})$, with weights
chosen empirically.
The soft Dice term is computed for the road class alone
using predicted probabilities, to account for the class
imbalance between road and non-road pixels.
The edge-weighted segmentation loss
$\mathcal{L}_{\mathrm{e\text{-}seg}}$ emphasizes track
boundary pixels by applying a per-pixel weight map
$\omega_i = 1 + 4 \cdot e_i$, where $e_i \in \{0,1\}$
is a binary boundary map derived from morphological erosion
of the ground-truth road labels, assigning a 5$\times$
weight to boundary pixels and unit weight to interior
pixels:
\begin{equation}
  \mathcal{L}_{\mathrm{e\text{-}seg}} =
  \frac{1}{2HW}\sum_{s \in \{L,R\}}\sum_{i=1}^{HW}
  \omega_i \cdot \mathcal{L}_{\mathrm{CE}}(\hat{y}_s(i),
  y_s(i))
  \label{eq:eseg_loss}
\end{equation}

\paragraph{ILC Loss}
While the geometry, photometry, and calibration losses
supervise inter-camera correspondence at the feature level,
the ILC loss ensures that correspondence is enforced
at the prediction level.
The ILC loss supervises the warp network by comparing
$\hat{y}^w_L$ against the left labels $y_L$:
\begin{equation}
  \mathcal{L}_{\mathrm{label}} =
  \mathcal{L}_{\mathrm{CE}}\!\left(\hat{y}^w_L,\; y_L\right)
  \label{eq:label_loss}
\end{equation}
While the inter-camera cross-attention mechanisms are
computed bidirectionally, the ILC module enforces geometric
consistency in one direction only. This is an empirical design choice.

\section{Training}
\label{sec:training}

\begin{figure}[t]
\centering
\includegraphics[width=\textwidth]{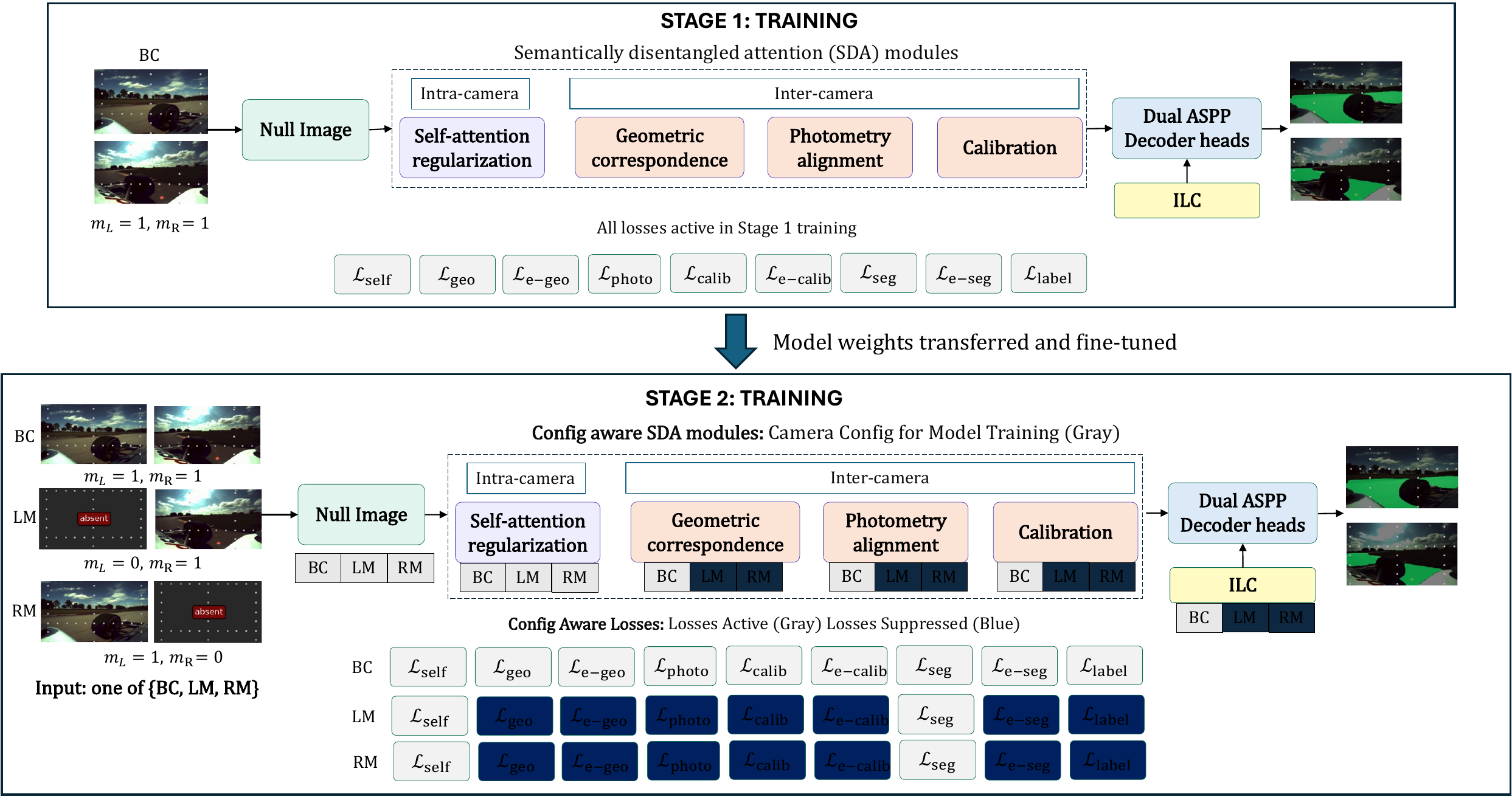}
\caption{Two-stage curriculum learning overview. Stage~1
trains on the BC partition and Stage~2 fine-tunes on
mixed-availability data~(BC~$\oplus$~LM~$\oplus$~RM).
The loss functions active for each sensor configuration
across both stages are illustrated.}
\label{fig:training}
\end{figure}

The proposed framework is trained using a two-stage curriculum
that accounts for the requirements of each module.
More specifically, the geometry, photometry, calibration, and
ILC modules require both cameras to be present, as their
losses depend on correspondence between the available camera images. Missing camera samples prevent the inter-camera attention mechanisms from effectively learning cross-camera representations.

To address this, a two-stage curriculum learning strategy is adopted. Stage~1 learns inter-camera
representations from fully-available BC data, and Stage~2
fine-tunes the trained framework on mixed-availability data
through config-aware loss masking.

\subsection{Stage 1: Multi-Camera Representation Learning}

In Stage~1, the proposed framework is trained on BC data
using all seven losses:
\begin{align}
  \mathcal{L}_{\mathrm{S1}} \;=\;&
    w_1 \mathcal{L}_{\mathrm{seg}}
  + w_2 \mathcal{L}_{\mathrm{e\text{-}seg}}
  + w_3 \mathcal{L}_{\mathrm{fusion}}
  \nonumber\\
  &+ w_4 \mathcal{L}_{\mathrm{geo}}
  + w_5 \mathcal{L}_{\mathrm{photo}}
  + w_6 \mathcal{L}_{\mathrm{calib}}
  + w_7 \mathcal{L}_{\mathrm{label}}
  \label{eq:stage1_loss}
\end{align}
with loss weights summarized in Table~\ref{tab:loss_weights}.

The full BC supervision ensures all modules are trained
with complete dual-camera data in Stage~1.
This allows the self-attention, cross-attention mechanism, and attention fusion mechanisms in the SDA
module, the ASPP decoder heads, and the ILC module to learn inter-camera
representations from fully-available paired data before
being exposed to missing camera configurations.
The learned weights are then fine-tuned in Stage~2 on
mixed-availability data through config-aware loss masking.

\begin{table}[ht]
\caption{Loss function weights used in training.
         Weights apply whenever the corresponding loss
         is active, as determined by the training stage
         and sensor configuration.}
\label{tab:loss_weights}
\centering
\begin{tabular}{lrrrrrrr}
\toprule
 & $\mathcal{L}_{\mathrm{seg}}$
 & $\mathcal{L}_{\mathrm{e\text{-}seg}}$
 & $\mathcal{L}_{\mathrm{fusion}}$
 & $\mathcal{L}_{\mathrm{geo}}$
 & $\mathcal{L}_{\mathrm{photo}}$
 & $\mathcal{L}_{\mathrm{calib}}$
 & $\mathcal{L}_{\mathrm{label}}$ \\
\midrule
$w_i$
 & 1.0 & 0.5 & 0.1 & 0.6 & 0.1 & 0.3 & 0.2 \\
\bottomrule
\end{tabular}
\end{table}

\subsection{Stage 2: Missing Camera Adaptation}
Stage~2 fine-tunes the Stage~1 pre-trained model on
mixed-availability data comprising BC, LM, and RM
configurations using config-aware loss masking.
No new losses are introduced, and the edge-weighted
segmentation loss $\mathcal{L}_{\mathrm{e\text{-}seg}}$
, as boundary emphasis was served in Stage~1. Stage~2 focuses on fine-tuning the weights to missing camera data through masking.

For BC samples, the Stage~2 loss comprises six active
losses:
\begin{align}
  \mathcal{L}_{\mathrm{S2}}^{\mathrm{BC}} \;=\;&
    w_1 \mathcal{L}_{\mathrm{seg}}
  + w_3 \mathcal{L}_{\mathrm{fusion}}
  + w_4 \mathcal{L}_{\mathrm{geo}}
  \nonumber\\
  &+ w_5 \mathcal{L}_{\mathrm{photo}}
  + w_6 \mathcal{L}_{\mathrm{calib}}
  + w_7 \mathcal{L}_{\mathrm{label}}
  \label{eq:stage2_bc_loss}
\end{align}

For LM samples~($m_L = 0$, $m_R = 1$), only the available
right camera stream contributes to the loss, as the absent
left camera provides no valid supervision signal:
\begin{equation}
  \mathcal{L}_{\mathrm{S2}}^{\mathrm{LM}} =
    w_1 \mathcal{L}_{\mathrm{seg}}(\hat{y}_R, y_R)
  + w_3 \mathcal{L}_{\mathrm{self}}(\mathbf{z}_R, \mathbf{t}_R)
  \label{eq:stage2_lm_loss}
\end{equation}

For RM samples~($m_L = 1$, $m_R = 0$), the symmetric case
applies, with only the available left camera stream
contributing:
\begin{equation}
  \mathcal{L}_{\mathrm{S2}}^{\mathrm{RM}} =
    w_1 \mathcal{L}_{\mathrm{seg}}(\hat{y}_L, y_L)
  + w_3 \mathcal{L}_{\mathrm{self}}(\mathbf{z}_L, \mathbf{t}_L)
  \label{eq:stage2_rm_loss}
\end{equation}

The loss weights are chosen empirically to balance the
contribution of each loss to the total gradient and are
not normalized. An overview of the curriculum learning is shown in Fig.~\ref{fig:training}.

\section{Inference}
At inference, the proposed framework processes input images
from one of three sensor configurations, BC, LM, or RM,
using the camera presence flags $(m_L, m_R)$.
The null image substitution module replaces the absent
camera input with the learned null image, providing a learnt representation to the feature extraction backbone.
The bidirectional cross-attention outputs are gated
independently by the presence flag of each key-value
camera. The dual ASPP decoder heads produce road segmentation predictions $\hat{y}_L$ and $\hat{y}_R$ for both camera views regardless of sensor configuration.
The ILC module and all auxiliary loss projections are
removed at inference, introducing no additional
computational overhead.

\section{Experiments}
\label{sec:experiments}

In this section, we validate the proposed RoadFormer-SDA framework using the RoRaTrack dataset~\cite{ghosh2025racingdatasetbaselinemodel}. A comparative analysis against baseline methods is reported, along with a model complexity analysis owing to low latency requirements for autonomous racing. Finally, a detailed ablation study is performed to validate the contribution of each proposed module.

\begin{table}[t]
\caption{Comparative analysis: proposed framework vs baseline
         algorithms~(B1--B4). Road~IoU~(\%).
         Avg~Avail~=~mean of available camera Road~IoU for LM
         and RM. Avg~Missing~=~mean of absent camera Road~IoU
         for LM and RM. Bold: best result per column.}
\label{tab:compar}
\centering
\begin{tabular}{lccc}
\toprule
Method & BC & Avg~Avail & Avg~Missing \\
\midrule
B1~\cite{Chen_2018_ECCV} & 92.80 & 93.32 & 18.31 \\
B2~\cite{Tan2019} & 86.67 & 87.45 & 18.31 \\
B3~\cite{wang2021_PAMI} & 88.28 & 88.45 & 18.31 \\
B4~\cite{Mehta_2022_MobileViT} & 87.12 & 87.79 & 18.31 \\
\midrule
Proposed  & \textbf{92.82} & \textbf{93.70} & \textbf{78.49} \\
\bottomrule
\end{tabular}
\end{table}

\begin{table}[t]
\caption{Comparative analysis: proposed framework vs baseline
         algorithms~(B1--B4). Avg~Avail~=~mean of
         available camera Road~IoU for LM and RM.
         Road~IoU~(\%). Bold: best result per column.}
\label{tab:compar_avail}
\centering
\begin{tabular}{lcccc}
\toprule
Method & BC & LM~Avail & RM~Avail & Avg~Avail \\
\midrule
B1~\cite{Chen_2018_ECCV} & 92.80 & 93.18 & 93.45 & 93.32 \\
B2~\cite{Tan2019} & 86.67 & 86.35 & 88.56 & 87.45 \\
B3~\cite{wang2021_PAMI} & 88.28 & 87.49 & 89.41 & 88.45 \\
B4~\cite{Mehta_2022_MobileViT} & 87.12 & 86.99 & 88.59 & 87.79 \\
\midrule
Proposed  & \textbf{92.82} & \textbf{93.54} & \textbf{93.85} & \textbf{93.70} \\
\bottomrule
\end{tabular}
\end{table}

\begin{table}[!h]
\caption{Comparative analysis: proposed framework vs baseline
         algorithms~(B1--B4). Avg~Missing~=~mean of
         absent camera Road~IoU for LM and RM.
         Road~IoU~(\%). Bold: best result per column.}
\label{tab:compar_missing}
\centering
\begin{tabular}{lccc}
\toprule
Method & LM~Missing & RM~Missing & Avg~Missing \\
\midrule
B1~\cite{Chen_2018_ECCV} & 20.50 & 16.12 & 18.31 \\
B2~\cite{Tan2019} & 20.50 & 16.12 & 18.31 \\
B3~\cite{wang2021_PAMI} & 20.50 & 16.12 & 18.31 \\
B4~\cite{Mehta_2022_MobileViT} & 20.50 & 16.12 & 18.31 \\
\midrule
Proposed  & \textbf{84.89} & \textbf{72.09} & \textbf{78.49} \\
\bottomrule
\end{tabular}
\end{table}

\subsection{Dataset}
\label{sec:experiments:dataset}

The proposed framework is validated using the RoRaTrack dataset~\cite{ghosh2025racingdatasetbaselinemodel}, a
dual-camera road segmentation dataset collected from an IAC autonomous racing vehicle equipped with two front-facing
cameras. The dataset comprises 606 paired frames captured across
multiple racing sessions. Each camera image has dimensions $180 \times 320$
pixels with 3 channels. The dataset is randomly partitioned into 70\%
training~(424 frames) and 30\% test~(182 frames) sets. 

To validate the proposed framework under missing modality
conditions, the original RoRaTrack dataset is extended to
simulate single camera failure. Each dual-camera image pair in the original dataset, together
with its corresponding dual ground truth maps, is termed the
\textit{Both Camera~(BC) partition}, with camera presence
flags $m_L = 1,\, m_R = 1$. The BC partition is duplicated twice to create two simulated
missing camera partitions. In the first duplicate, the left camera image is set to
absent to create the \textit{Left Missing~(LM) partition}
with camera presence flags $m_L = 0,\, m_R = 1$, while the
right camera image and dual ground truth maps are retained
unchanged.
In the second duplicate, the right camera image is set to
absent to create the \textit{Right Missing~(RM) partition}
with camera presence flags $m_L = 1,\, m_R = 0$, while the
left camera image and dual ground truth maps are retained
unchanged.

The combined training split therefore comprises 1,272
mixed-availability samples~(424 BC, 424 LM, 424 RM), and
the test split comprises 546 samples~(182 BC, 182 LM,
182 RM). The camera presence flags $(m_L, m_R)$ are provided per
sample for both training and evaluation.

Stage~1 of the proposed framework is trained on the BC partition
only, while Stage~2 of the proposed framework and all
baseline algorithms are trained on the full
mixed-availability training split comprising BC, LM, and
RM partitions. The inference for the proposed, baseline, and ablated frameworks are performed on the mixed-availability test split.

\subsection{Algorithm Parameters}
\label{sec:experiments:params}

Given the inputs $\mathbf{I}_L$, $\mathbf{I}_R$, the null image module generates two learnable null image
tensors $\mathbf{N}_s \in \mathbb{R}^{3 \times 180 \times 320}$,
one per camera, initialized to the ImageNet channel
mean and updated during training. The feature extraction backbone uses the
DeepLabV3$+$ResNet-50 architecture~\cite{Chen_2018_ECCV}, pretrained on ImageNet, to extract the features independently from $\mathbf{N}_s$. The features extracted are sent to the SDA module, which uses an attention head dimension of $d = C/4$
with a spatial average pooling factor of $p = 2$ applied
before the attention operations. Each ASPP decoder head applies dilated convolutions at rates
$\{1, 6, 12\}$ with a $1 \times 1$ fusion convolution and
a dropout rate of 0.1. The ILC warp network comprises two convolutional layers with
32 intermediate channels, producing a dense flow field
$\phi \in \mathbb{R}^{H \times W \times 2}$.

\subsection{Baseline Algorithms}
\label{sec:experiments:baselines}

\begin{figure}[t]
\centering
\includegraphics[width=\textwidth]{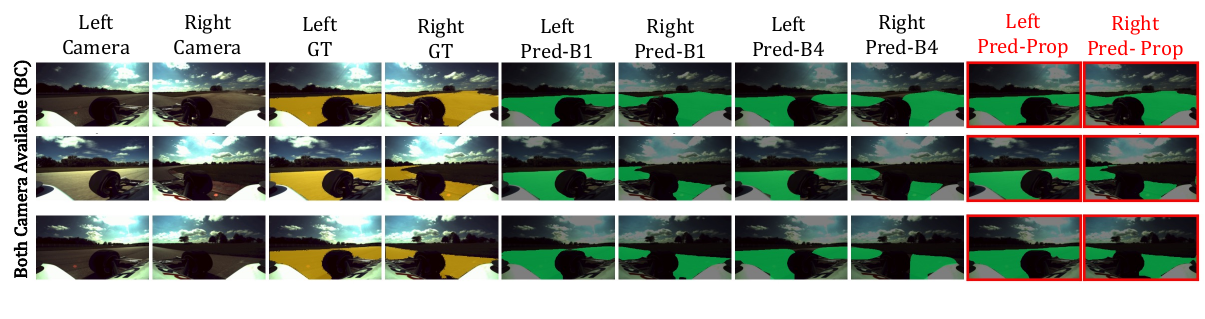}
\caption{Qualitative comparison for the BC partition~(both cameras available). Columns show left and right camera inputs, ground truth masks, baseline predictions~(B1 and B4), and proposed framework predictions~(red border)}
\label{fig:result_bc}
\end{figure}

\begin{figure}[t]
\centering
\includegraphics[width=\textwidth]{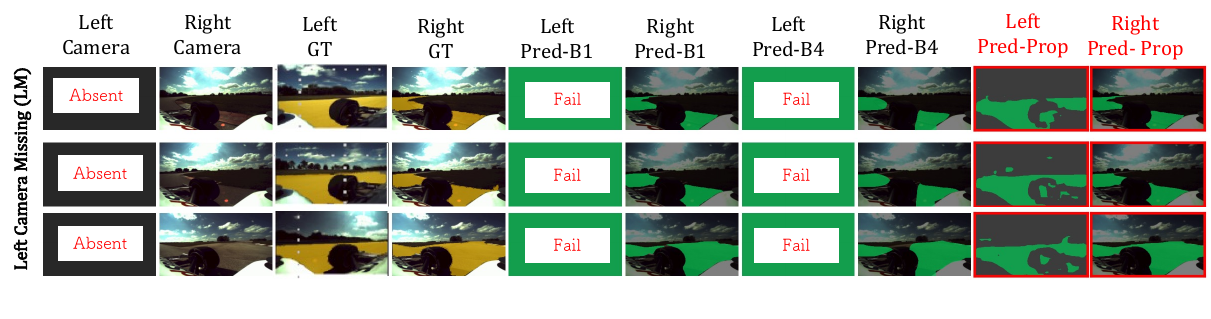}
\caption{Qualitative comparison for the LM configuration~(left camera absent). Baseline models fail to produce meaningful left camera predictions~(Fail), while the proposed framework estimates the free space.}
\label{fig:result_lm}
\end{figure}

\begin{figure}[t]
\centering
\includegraphics[width=\textwidth]{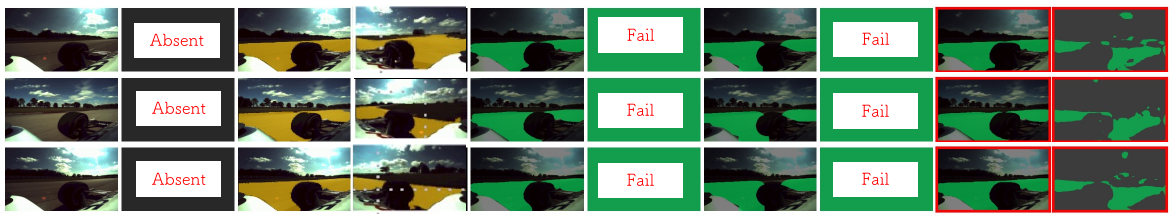}
\caption{Qualitative comparison for the RM configuration~(right camera absent). Baseline models fail to produce  right camera predictions~(Fail), while the proposed framework estimates the free space.}
\label{fig:result_rm}
\end{figure}

The first baseline uses the DeepLabV3$+$ResNet-50
architecture~\cite{Chen_2018_ECCV}. Hereafter, this baseline is referred to as B1. The second baseline uses the EfficientNet-B2
architecture~\cite{Tan2019}. Hereafter, this baseline is referred to as B2. The third baseline uses the HRNet-W18
architecture~\cite{wang2021_PAMI}. Hereafter, this baseline is referred to as B3. The fourth baseline uses the MobileViT-XXS
architecture~\cite{Mehta_2022_MobileViT}. Hereafter, this baseline is referred to as B4.

Each baseline is trained with the mixed-availability dataset, and using zero substitution for the missing camera input. All baselines use independent feature extraction modules for each camera input, followed by a simple addition fusion of left and right backbone features, dual ASPP decoder heads, and the combined CE+Dice segmentation loss.

\subsection{Training Details and Evaluation Metric}
\label{sec:experiments:params}

All models are implemented in PyTorch and trained on an NVIDIA
RTX~5090 with backbone weights initialized from ImageNet
pretraining. Stage~1 uses AdamW with learning rate $3 \times 10^{-4}$, weight
decay $10^{-4}$, cosine annealing schedule, and a 5-epoch linear
warmup for 60 epochs at batch size~16. Stage~2 uses AdamW with learning rate $1 \times 10^{-4}$, cosine annealing schedule, and a 3-epoch linear warmup for 60 epochs. The loss weights for both stages are summarized in
Table~\ref{tab:loss_weights}. The training parameters are empirically selected and represent the best performances of the different algorithms.

\textit{Evaluation Metric.} Road~IoU~(class~0 intersection-over-union) is adopted as the primary evaluation metric to account for the  class imbalance:
\begin{equation}
  \text{Road~IoU} =
  \frac{TP_{\mathrm{road}}}
       {TP_{\mathrm{road}} + FP_{\mathrm{road}} + FN_{\mathrm{road}}}
  \label{eq:road_iou}
\end{equation}

For missing camera configurations, two metrics are reported. The average available camera Road~IoU is defined as the mean Road~IoU of the available camera head across LM and RM configurations. More specifically, the right head for LM and the left head
for RM. The average missing camera Road~IoU is defined as the mean
Road~IoU of the absent camera head across LM and RM
configurations, the left head for LM and the right head
for RM. Here, predictions for the missing cameras are evaluated against the ground truth labels retained in the test split. The average missing camera Road~IoU measures the framework's ability to generate road segmentation for the absent camera view at testing.

Finally, as real-time operation with low latency and memory constraints are critical for deployment on edge devices in autonomous racing, the inference latency and memory are reported for one image pair with FP16 precision.

\subsection{Comparative Analysis}
\label{sec:compar}

The performance of the proposed framework and the baseline
algorithms, all trained on mixed-availability data, are reported in Table~\ref{tab:compar}, Table~\ref{tab:compar_avail}, and Table~\ref{tab:compar_missing}.

The results show that the proposed framework reports better
average available camera Road~IoU than the baseline algorithms,
and substantially better average missing camera Road~IoU across
all missing modality configurations. A few examples of the estimated semantic segmentation for the varying sensor configurations are shown in Fig.~\ref{fig:result_bc}-Fig.~\ref{fig:result_rm}.

While the baseline algorithms report varying average available camera Road~IoU, the average missing camera Road~IoU is identical at 18.31\% for all four baselines, indicating that the absent camera prediction fails
consistently regardless of the baseline architecture.
This confirms that zero substitution and simple addition-based fusion is insufficient for absent camera prediction. In contrast, the proposed framework achieves 78.49\% average missing camera Road~IoU.

\subsection{Ablation Study}

\begin{table}[t]
\caption{Ablation configuration summary. For each ablation, the removed attention mechanism, fusion status, and removed loss are listed.
         \checkmark~=~present. $\times$~=~removed.}
\label{tab:ablation_config}
\centering
\begin{tabular}{lcccc}
\toprule
Configuration & Attention & Fusion & Loss Removed \\
\midrule
Proposed      & \checkmark & \checkmark & -- \\
\midrule
A1: No geometry    & $\times$ geo   & \checkmark & $\mathcal{L}_{\mathrm{geo}}$ \\
A2: No calibration & $\times$ calib & \checkmark & $\mathcal{L}_{\mathrm{calib}}$ \\
A3: No photometry  & $\times$ photo & \checkmark & $\mathcal{L}_{\mathrm{photo}}$ \\
A4: No null image  & \checkmark     & \checkmark & -- \\
A5: No ILC module  & \checkmark     & \checkmark & $\mathcal{L}_{\mathrm{label}}$ \\
A6: No fusion loss & \checkmark     & \checkmark & $\mathcal{L}_{\mathrm{fusion}}$ \\
\bottomrule
\end{tabular}
\end{table}

An ablation study is conducted to validate the contribution of each proposed component to the final road segmentation performance. By ablating different components in the proposed framework, we validate the importance of the different components. A complete list of ablated is shown in Table~\ref{tab:ablation_config}. For A1--A5, both the module and its corresponding loss are removed from the architecture across Stage~1 and Stage~2
training. For A6, only the fusion loss $\mathcal{L}_{\mathrm{fusion}}$ is removed. The attention fusion output $\mathbf{z}_s$ is
retained but without the constraint to remain close to the input token sequences $\mathbf{t}_s$.
The performance of the proposed configuration and each
ablation variant are reported in
Table~\ref{tab:detailed_ablation}.

\noindent\textit{Available Cameras in Missing Modality
Configuration.}
All ablation variants report lower Avg~Avail than the
proposed configuration, confirming that every component
contributes positively to present camera performance.
The omission of calibration loss reports the largest
difference~($-$3.06\%), followed by the omission of the photometry and the ILC losses, ~($-$2.40\%) and ~($-$2.14\%), respectively. The null image,~ the geometry and fusion loss omissions report the smallest differences of ($-$2.10\%), $-$1.91\% and $-$1.82\% respectively.

\noindent\textit{Missing Cameras in Missing Modality
Configuration.}
All ablation variants also report lower Avg~Missing than the proposed configuration, confirming that every component contributes positively to absent camera prediction quality. The omission of photometry loss reports the largest difference~($-$23.65\%), followed by the omission of fusion loss~($-$21.90\%), the omission of ILC module~($-$14.38\%), and the omission of calibration loss~($-$10.10\%).
The omission of null image reports a difference of $-$4.98\%, while the
omission of geometry loss reports the smallest difference of $-$0.94\%.

\section{Discussion}
\label{sec:discussion}

\begin{table}[t]
\caption{Ablation on loss components. 
         Road~IoU~(\%).
         Avg~Avail~=~mean of available camera Road~IoU for LM
         and RM. Avg~Missing~=~mean of absent camera Road~IoU
         for LM and RM. $\Delta$~=~difference from Proposed.}
\label{tab:detailed_ablation}
\centering
\begin{tabular}{lccccc}
\toprule
Configuration & BC & Avg~Avail & $\Delta$ & Avg~Missing & $\Delta$ \\
\midrule
Proposed              & \textbf{92.82} & \textbf{93.70} & --       & \textbf{78.49} & --       \\
\midrule
A1: No geometry       & 91.02 & 91.79 & $-$1.91  & 77.56 & $-$0.94  \\
A2: No calibration    & 91.55 & 90.64 & $-$3.06  & 68.39 & $-$10.10 \\
A3: No photometry     & 90.84 & 91.29 & $-$2.40  & 54.85 & $-$23.65 \\
A4: No null image     & 91.90 & 91.60 & $-$2.10  & 73.51 & $-$4.98  \\
A5: No ILC head       & 90.40 & 91.55 & $-$2.14  & 64.12 & $-$14.38 \\
A6: No Fusion Loss. & 91.35 & 91.87 & $-$1.82  & 56.59 & $-$21.90 \\
\bottomrule
\end{tabular}
\end{table}

\noindent \textbf{Comparative Analysis.}
The results in Tables~\ref{tab:compar},~\ref{tab:compar_avail},
and~\ref{tab:compar_missing} demonstrate that the proposed
framework reports better average present camera Road~IoU
than all four baseline algorithms.
More importantly, as reported in Table~\ref{tab:compar_missing},
the proposed framework substantially improves average absent
camera Road~IoU compared to all four baselines, which report
a uniform 18.31\% corresponding to degenerate road-class
frequency prediction from zero input.
This can be attributed to the semantically disentangled
attention mechanism supervised by seven disentangled loss
functions, complemented by the learned null image, which
provides structured missing camera representations in place
of naive zero substitution.

\noindent \textbf{Curriculum Learning.}
While the baseline algorithms have no restriction on the
type of training data and can be directly trained on
mixed-availability data, the proposed framework requires
a two-stage curriculum.
Stage~1 is trained on BC-only data to learn structured
inter-camera representations, as jointly training on
mixed-availability data would deprive the inter-camera
modules of supervision from missing camera samples.
Stage~2 then fine-tunes on mixed-availability data to
adapt to missing camera configurations through config-aware
loss masking.

\noindent \textbf{Ablation on Loss Components.}
The results in Table~\ref{tab:detailed_ablation} show that all components contribute positively to both Avg~Avail and Avg~Missing, confirming that the full proposed configuration is necessary for optimal performance on both metrics.

For Avg~Avail, the calibration loss reports the largest
contribution, which is consistent with the importance of inter-camera extrinsic alignment in autonomous racing where mechanical stress and vibration can induce calibration drift. The photometry and ILC losses are the next most critical components. The photometry loss reflects the importance
of cross-camera appearance alignment, while the ILC loss enforces geometric consistency between the left and right segmentation predictions at the prediction level. The geometry and fusion losses report the smallest contributions to Avg~Avail.

For Avg~Missing, the photometry and fusion losses are the most critical components. The photometry loss aligns cross-camera appearance representations during BC training in Stage-1 and Stage-2. The learned alignment from BC training transfers to missing camera configurations at inference, where the available camera stream queries the null image tokens of the absent camera.

The fusion loss constrains the per-camera fused
representation to remain close to the input tokens.
For missing camera configurations, only the available
camera stream contributes to $\mathbf{F}^{\mathrm{fused}}$, so the fusion loss is critical for the available stream
$\mathbf{z}_s$. This ensures the attention fusion retains a compact representation of the input tokens corresponding to the available camera. 

The ILC and calibration losses also contribute, as geometric consistency and extrinsic
offset modeling learned from BC samples directly support road segmentation inference from the available camera  under missing camera configurations.
The null image contributes meaningfully to Avg~Missing,
as providing an in-distribution input to the backbone for the absent camera is necessary for
feature extraction. The geometry loss reports the smallest contribution to Avg~Missing.

The proposed configuration reports the best performance
on both Avg~Avail and Avg~Missing across all ablation
variants.

\noindent \textbf{Deployment.}
The proposed framework has a memory requirement of 283~MB
and an inference latency of 4.51~ms at batch size~1 with
FP16 precision, which is within the real-time constraints
of the autonomous racing platform.

\noindent \textbf{Limitations and Future Work.}
As part of our future work, we will investigate the
generalization of the proposed framework to additional
stereo road segmentation datasets beyond RoRaTrack to
validate performance across different camera configurations
and racing environments.
The binary camera presence flag~($m_s \in \{0, 1\}$) does
not model partial degradation scenarios such as lens
contamination or image saturation.
In future work, we will investigate varying degrees of
camera degradation to handle scenarios beyond complete
camera failure.

\section{Conclusions}
\label{sec:conclusion}

In this paper, we propose a multi-camera road segmentation
framework for autonomous racing under full and partial sensor
availability.
The framework comprises a learned null image substitution,
a semantically disentangled attention~(SDA) module with five
independent attention mechanisms supervised by multiple
disentangled loss functions, dual ASPP decoder heads, and
an inter-camera label consistency head.
The framework is trained using a two-stage curriculum that
first establishes structured inter-camera representations on
fully-available data, then adapts to missing camera
conditions.
A comparative analysis on the RoRaTrack dataset with baseline
methods validates the proposed framework, reporting consistent
improvement over the baseline.
The ablation study confirm the importance of each module in the proposed framework.
As part of our future work, we will investigate the
generalization of the proposed framework to additional stereo
road segmentation datasets, and explore continuous camera
quality modeling beyond the binary presence flag.

\bibliographystyle{unsrt} 
\bibliography{references}

@INPROCEEDINGS{Cordts_2016,
  author={Cordts, Marius and Omran, Mohamed and Ramos, Sebastian and Rehfeld, Timo and Enzweiler, Markus and Benenson, Rodrigo and Franke, Uwe and Roth, Stefan and Schiele, Bernt},
  booktitle={IEEE Conference on Computer Vision and Pattern Recognition (CVPR)}, 
  title={The Cityscapes Dataset for Semantic Urban Scene Understanding}, 
  year={2016},
  pages={3213-3223},
}

@misc{ghosh2025racingdatasetbaselinemodel,
      title={A Racing Dataset and Baseline Model for Track Detection in Autonomous Racing}, 
      author={Shreya Ghosh and Yi-Huan Chen and Ching-Hsiang Huang and Abu Shafin Mohammad Mahdee Jameel and Chien Chou Ho and Aly El Gamal and Samuel Labi},
      year={2025},
      eprint={2502.14068},
      archivePrefix={arXiv},
      primaryClass={cs.CV},
}

@INPROCEEDINGS{Neuhold_2017,
  author={Neuhold, Gerhard and Ollmann, Tobias and Bulò, Samuel Rota and Kontschieder, Peter},
  booktitle={2017 IEEE International Conference on Computer Vision (ICCV)}, 
  title={The Mapillary Vistas Dataset for Semantic Understanding of Street Scenes}, 
  year={2017},
  volume={},
  number={},
  pages={5000-5009},
}

@article{Betz_2022,
 title = {Autonomous Vehicles on the Edge: A Survey on Autonomous Vehicle Racing},
 journal = {IEEE Open Journal of Intelligent Transportation Systems}, 
author = {Betz, Johannes and Zheng, Hongrui and Liniger, Alexander and Rosolia, Ugo and Karle, Phillip and Behl, Madhur and Krovi, Venkat and Mangharam, Rahul},
volume = {3},
pages ={458--488},
year={2022}
}

@article{Betz_2023,
author = {Betz, Johannes and Betz, Tobias and Fent, Felix and Geisslinger, Maximilian and Heilmeier, Alexander and Hermansdorfer, Leonhard and Herrmann, Thomas and Huch, Sebastian and Karle, Phillip and Lienkamp, Markus and Lohmann, Boris and Nobis, Felix and Ögretmen, Levent and Rowold, Matthias and Sauerbeck, Florian and Stahl, Tim and Trauth, Rainer and Werner, Frederik and Wischnewski, Alexander},
title = {TUM autonomous motorsport: An autonomous racing software for the Indy Autonomous Challenge},
journal = {Journal of Field Robotics},
volume = {40},
number = {4},
pages = {783--809},
year = {2023},
}

@inproceedings{Long_FCN,
  author={Long, Jonathan and Shelhamer, Evan and Darrell, Trevor},
  booktitle={2015 IEEE Conference on Computer Vision and Pattern Recognition (CVPR)}, 
  title={Fully convolutional networks for semantic segmentation}, 
  year={2015},
  pages={3431--3440},
}

@inproceedings{Chen_2018_ECCV,
author = {Chen, Liang-Chieh and Zhu, Yukun and Papandreou, George and Schroff, Florian and Adam, Hartwig},
title = {Encoder-Decoder with Atrous Separable Convolution for Semantic Image Segmentation},
booktitle = {Proceedings of the European Conference on Computer Vision (ECCV)},
month = {September},
year = {2018}
}

@article{wang2021_PAMI,
author = {Wang, Jingdong and Sun, Ke and Cheng, Tianheng and Jiang, Borui and Deng, Chaorui and Zhao, Yang and Liu, Dong and Mu, Yadong and Tan, Mingkui and Wang, Xinggang and Liu, Wenyu and Xiao, Bin},
title = {Deep High-Resolution Representation Learning for Visual Recognition},
year = {2021},
month = {September},
pages = {3349--3364},
journal = {IEEE Transactions on Pattern Analysis and Machine Intelligence},
volume = {43},
number = {10},
}

@inproceedings{Xie_2021_Segformer,
author = {Xie, Enze and Wang, Wenhai and Yu, Zhiding and Anandkumar, Anima and Alvarez, Jose M. and Luo, Ping},
title = {SegFormer: simple and efficient design for semantic segmentation with transformers},
year = {2021},
booktitle = {Proceedings of the 35th International Conference on Neural Information Processing Systems},
articleno = {924},
numpages = {14},
series = {NIPS '21}
}

@InProceedings{Tan2019,
  title = 	 {{E}fficient{N}et: Rethinking Model Scaling for Convolutional Neural Networks},
  author =       {Tan, Mingxing and Le, Quoc},
  booktitle = 	 {Proceedings of the International Conference on Machine Learning},
  pages = 	 {6105--6114},
  year = 	 {2019},
}

@misc{jaderberg2015spatial,
      title={Spatial Transformer Networks}, 
      author={Max Jaderberg and Karen Simonyan and Andrew Zisserman and Koray Kavukcuoglu},
      year={2016},
      eprint={1506.02025},
      archivePrefix={arXiv},
      primaryClass={cs.CV},
}

@misc{Mehta_2022_MobileViT,
      title={Separable Self-attention for Mobile Vision Transformers}, 
      author={Sachin Mehta and Mohammad Rastegari},
      year={2022},
      eprint={2206.02680},
      archivePrefix={arXiv},
      primaryClass={cs.CV},
}

@InProceedings{valada16iser,
author = {Abhinav Valada and Gabriel Oliveira and Thomas Brox and Wolfram Burgard},
title = {Deep Multispectral Semantic Scene Understanding of Forested Environments using Multimodal Fusion},
booktitle = {International Symposium on Experimental Robotics (ISER)},
year = {2016},
}

@inproceedings{vaswani2017attention,
  author    = {Vaswani, Ashish and Shazeer, Noam and Parmar, Niki and Uszkoreit, Jakob and Jones, Llion and Gomez, Aidan N. and Kaiser, {\L}ukasz and Polosukhin, Illia},
  title     = {Attention is All You Need},
  booktitle = {Advances in Neural Information Processing Systems},
  volume    = {30},
  pages     = {5998--6008},
  year      = {2017},
}

@inproceedings{CarionDETR,
author = {Carion, Nicolas and Massa, Francisco and Synnaeve, Gabriel and Usunier, Nicolas and Kirillov, Alexander and Zagoruyko, Sergey},
title = {End-to-End Object Detection with Transformers},
year = {2020},
booktitle = {European Conference Computer Vision},
pages = {213–229},
}

@inproceedings{Li_BEVFormer,
author = {Li, Zhiqi and Wang, Wenhai and Li, Hongyang and Xie, Enze and Sima, Chonghao and Lu, Tong and Qiao, Yu and Dai, Jifeng},
title = {BEVFormer: Learning Bird’s-Eye-View Representation from Multi-camera Images via Spatiotemporal Transformers},
year = {2022},
booktitle = {European Conference Computer Vision},
pages = {1–18},
}

@misc{chitta2022transfuserimitationtransformerbasedsensor,
      title={TransFuser: Imitation with Transformer-Based Sensor Fusion for Autonomous Driving}, 
      author={Kashyap Chitta and Aditya Prakash and Bernhard Jaeger and Zehao Yu and Katrin Renz and Andreas Geiger},
      year={2022},
      eprint={2205.15997},
      archivePrefix={arXiv},
      primaryClass={cs.CV},
}

@inproceedings{Xiong_2025,
author = {Xiong, Yuzhu and Du, Hongwei and Zhang, Daikun},
title = {A Guided-learning Network for Brain Tumor Segmentation with Missing Modalities},
year = {2025},
booktitle = {Proceedings of the International Conference on Generative Artificial Intelligence and Information Security},
pages = {219–225},
numpages = {7},
}

@misc{havaei2016hemisheteromodalimagesegmentation,
      title={HeMIS: Hetero-Modal Image Segmentation}, 
      author={Mohammad Havaei and Nicolas Guizard and Nicolas Chapados and Yoshua Bengio},
      year={2016},
      eprint={1607.05194},
      archivePrefix={arXiv},
      primaryClass={cs.CV},
      url={https://arxiv.org/abs/1607.05194}, 
}

@misc{wu2026deepmultimodallearningmissing,
      title={Deep Multimodal Learning with Missing Modality: A Survey}, 
      author={Renjie Wu and Hu Wang and Hsiang-Ting Chen and Gustavo Carneiro},
      year={2026},
      eprint={2409.07825},
      archivePrefix={arXiv},
      primaryClass={cs.CV},
      url={https://arxiv.org/abs/2409.07825}, 
}

@inproceedings{Hussen2020,
author = {Hussen Abdelaziz, Ahmed and Theobald, Barry-John and Dixon, Paul and Knothe, Reinhard and Apostoloff, Nicholas and Kajareker, Sachin},
title = {Modality Dropout for Improved Performance-driven Talking Faces},
year = {2020},
booktitle = {Proceedings of the 2020 International Conference on Multimodal Interaction},
pages = {378–386},
numpages = {9},
}

@inproceedings{recasens2023zorro,
  title     = {Zorro: the masked multimodal transformer},
  author    = {Recasens, Adri\`a and Sun, Chen and Ross, David A. and Vondrick, Carl and Zisserman, Andrew},
  booktitle = {Proceedings of the IEEE/CVF Conference on Computer Vision and Pattern Recognition},
  pages     = {2995--3005},
  year      = {2023}
}

@article{JohnTOMM2025,
author = {John, Vijay and Kawanishi, Yasutomo},
title = {Multimodal Cascaded Framework with Multimodal Latent Loss Functions Robust to Missing Modalities},
year = {2025},
volume = {21},
number = {9},
journal = {ACM Trans. Multimedia Comput. Commun. Appl.},
month = sep,
articleno = {255},
numpages = {21},
}

@inproceedings{john2023multimodal,
  title={Multimodal cascaded framework with metric learning robust to missing modalities for person classification},
  author={John, Vijay and Kawanishi, Yasutomo},
  booktitle={Proceedings of the 14th Conference on ACM Multimedia Systems},
  pages={257--265},
  year={2023},
}

@inproceedings{john2023audiovisual,
  title={Audio-visual sensor fusion framework using person attributes robust to missing visual modality for person recognition},
  author={John, Vijay and Kawanishi, Yasutomo},
  booktitle={International Conference on Multimedia Modeling},
  pages={523--535},
  year={2023},
}

@inproceedings{john2022multimodal,
  title={A multimodal sensor fusion framework robust to missing modalities for person recognition},
  author={John, Vijay and Kawanishi, Yasutomo},
  booktitle={Proceedings of the 4th ACM International Conference on Multimedia in Asia},
  pages={1--5},
  year={2022},
}

@inproceedings{bengio_curriculum_2009,
	title = {Curriculum learning},
	booktitle = {Proceedings of the {International} {Conference} on {Machine} {Learning}},
	author = {Bengio, Yoshua and Louradour, Jérôme and Collobert, Ronan and Weston, Jason},
	year = {2009},
	pages = {1--8},
}

@inproceedings{lin2017focal,
  title     = {Focal Loss for Dense Object Detection},
  author    = {Lin, Tsung-Yi and Goyal, Priya and Girshick, Ross B. and He, Kaiming and Doll{\'{a}}r, Piotr},
  booktitle = {Proceedings of the IEEE International Conference on Computer Vision (ICCV)},
  year      = {2017},
  pages     = {2999--3007},
}

@inproceedings{godard2019monodepth2,
  title     = {Digging Into Self-Supervised Monocular Depth Estimation},
  author    = {Godard, Cl{\'e}ment and {Mac Aodha}, Oisin and Firman, Michael and Brostow, Gabriel J.},
  booktitle = {Proceedings of the IEEE/CVF International Conference on Computer Vision (ICCV)},
  pages     = {3828--3838},
  year      = {2019}
}

@inproceedings{zhou2017unsupervised,
  title     = {Unsupervised Learning of Depth and Ego-Motion from Video},
  author    = {Zhou, Tinghui and Brown, Matthew and Snavely, Noah and Lowe, David G.},
  booktitle = {Proceedings of the IEEE Conference on Computer Vision and Pattern Recognition (CVPR)},
  pages     = {6612--6619},
  year      = {2017}
}

@misc{zhao2026crossviewdomaingeneralizationgeometric,
      title={Cross-view Domain Generalization via Geometric Consistency for LiDAR Semantic Segmentation}, 
      author={Jindong Zhao and Yuan Gao and Yang Xia and Sheng Nie and Jun Yue and Weiwei Sun and Shaobo Xia},
      year={2026},
      eprint={2602.14525},
      archivePrefix={arXiv},
      primaryClass={cs.CV},
}

\end{document}